\documentclass[aps,prb,showpacs,preprint]{revtex4}
\usepackage{setspace}
\usepackage{mathrsfs}
\usepackage{graphicx}
\usepackage{bm}
\usepackage{amsmath,amssymb}
\usepackage{subfigure}

\begin{document}

\title{Effect of polydispersity on the relative stability of hard-sphere crystals}

\author{ Mingcheng Yang}
\author{Hongru Ma}
\email{hrma@sjtu.edu.cn} \affiliation{Institute of Theoretical
Physics, Shanghai Jiao Tong University,
 Shanghai 200240, People's Republic of China}
\date{\today}

\begin{abstract}
By extending the nonequilibrium potential refinement algorithm(NEPR)
and lattice switch(LW) method to the semigrand ensemble, the
semigrand potentials of the $fcc$ and $hcp$ structure of
polydisperse hard sphere crystals are calculated with the bias
sampling scheme. The result shows that the $fcc$ structure is more
stable than the $hcp$ structure for polydisperse hard sphere
crystals below the terminal polydispersity.

\end{abstract}


\maketitle

\section{Introduction} \label{intr}

A set of hard spheres under thermal agitation constitute a simple
yet non trivial model of condensed matter and especially represents
an idealization of a very important class of real colloid
dispersions. The model has been extensively studied in the past
decades. One of the important feature of the model is that the
system undergoes a purely entropy-driven first-order phase
transition from the fluid phase to a crystal phase at sufficiently
high density\cite{T. E. Wainwright,W.W. Wood,W.G. Hoover}. Simple
estimations of the free energy of different crystal structures
reveal that the possible structure could be  face-centered cubic
($fcc$) or hexagonal close packed ($hcp$). However, due to the very
similarity in local environments of the two structures, the
difference of free energy between them is extremely small and very
hard to determine. The determination of the relative stability
between the two structures from theoretical calculations has a long
history\cite{B. J. Alder}. A clear consensus was reached in the last
decade that $fcc$ is the more stable phase\cite{L. V. Woodcock,P.G.
Bolhuis1,A. D. Bruce}, however, there are still different views on
the problem. The recent results of Pronk and Frenkel \cite{S. Pronk}
indicate that a moderate deformation of a hard-sphere crystal may
make the $hcp$ phase more stable than the $fcc$ phase.  Kwak and
Kofke \cite{S. K. Kwak} investigated the effect of monovacancies on
the relative stability of $fcc$ and $hcp$ hard-sphere crystals.

The particle size of most artificial colloidal systems are
polydisperse, the polydispersity is defined as the ratio of the
standard deviation and the mean of the diameter distribution of
particles, which is an intrinsic property of a colloidal system. The
polydispersity may significantly affect the thermodynamic and
dynamic properties of a hard sphere system, e.g., there exists a
terminal polydispersity above which no cystallization can
occur\cite{P. N. Pusey,P.G. Bolhuis,M. Fasolo}, the osmotic pressure
of a polydisperse hard sphere crystal is higher than the one of a
monodisperse system with the same volume fraction\cite{S. Phan,S.
Phan1}, and there are local fractionations of particle sizes which
has a strong retarding effect on nucleation\cite{S. Martin,H. J.
Schope}. However, the influence of size polydispersity on the
relative stability of $fcc$ and $hcp$ hard-sphere crystals has not
been addressed. In this paper we will compute the free energy
difference between polydisperse $fcc$ and $hcp$ hard sphere crystals
by Monte Carlo(MC) simulations, which suggests that the $fcc$ phase
is still more stable than $hcp$ phase below the terminal
polydispersity.

The simple simulating method in canonical ensemble is not suitable
for a polydisperse system with  continuous distribution of particle
sizes. The reason is very simple,  the simulation system is often
too small to realize a given particle size distribution to the
designed accuracy.  Thus the grand canonical ensemble or
semigrand\cite{J. G. Briano} ensemble must be used. In these
ensembles the number of particles of each size (characterized by the
particle diameter $\sigma$), $N(\sigma)$ (thus the size distribution
$P(\sigma)$) is permitted to fluctuate, therefore, they can simulate
a true polydisperse system in the average sense. Comparing with the
grand canonical ensemble, the semigrand ensemble is more suitable
because the total number of particles is fixed and the insertion or
deletion of particles is not needed. The semigrand ensemble is
especially more suited to simulate the dense fluid and
crystal\cite{D. A. Kofke2,M. A. Bates,P.G. Bolhuis}, which provides
us a perfect framework to investigate the stability of polydisperse
hard sphere crystals.

In the semigrand ensemble, the particle size distribution $P(\sigma)$
is not chosen \emph{a priori}, which is obtained only after the simulation
 has been performed\cite{D. A.
Kofke2,M. A. Bates,P.G. Bolhuis}. This is because the imposed
physical variables in the simulation are not the composition
distribution but the chemical potential deference function
$\Delta\mu(\sigma)=\mu(\sigma)-\mu(\sigma_{r})$ (here $\sigma_{r}$
is the diameter of an arbitrarily chosen reference component), which
is a functional of the composition distribution $P(\sigma)$.
Consequently, in order to simulate a system with a prescribed
distribution  the inverse problem
$\Delta\mu(\sigma)=\Delta\mu(\{P(\sigma)\})$ has to be solved.
Recently, Escobedo\cite{F. Escobedo} and Wilding et al\cite{N. B.
Wilding} have separately shown that the inverse problem can be
solved by a histogram reweighting method. Alternatively, a more
robust and convenient scheme, the so called nonequilibrium potential
refinement algorithm(NEPR), was proposed by Wilding\cite{N. B.
Wilding1} and works excellent in the grand canonical simulation. We
will extend the algorithm to the semigrand ensemble, and  use the
extended method to determine the chemical potential deference
function $\Delta\mu(\{P(\sigma)\})$  of an arbitrarily prescribed
composition distribution $P(\sigma)$ in a semigrand ensemble. The
resulting forms of $\Delta\mu(\{P(\sigma)\})$ are then used to study
the stability of the polydisperse hard sphere crystals.

To avoid any confusion with the Helmholtz free energy, we will refer
to the free energy of the semigrand ensemble as the semigrand free
energy. In the semigrand ensemble the most stable phase has the
lowest semigrand free energy.  There are basically two routes  to
follow in the evaluation of the semigrand free energy. One is the
thermodynamic integration route\cite{D. Frenkel,D. Frenkel1} which
determines the free energy of a system  by integrating its
derivatives along a parameter space path connecting the system of
interest to a reference system( e.g., Einstein solid or ideal gas).
The other is the lattice switch method proposed by Bruce et
al\cite{A. D. Bruce}, from which the free energy difference between
monodisperse $fcc$ and $hcp$ hard-sphere crystals can be calculated
more directly than the thermodynamic integration method. The method
is utilized in the canonical\cite{A. D. Bruce} and
isobaric-isothermal ensemble\cite{N. B. Wilding2}. Therefore, We
will extend the lattice switch approach to the semigrand ensemble in
the present work and use it for the study of thermodynamic stability
of the polydisperse hard sphere system.

The contents of the remain of this paper are as follows. In section
\ref{semi_ens} we formulate the statistical mechanics for a
polydisperse system within the semigrand ensemble. In section
\ref{method} the methodology employed in the work is described and
their validity is checked. The computational details and results are
presented in section \ref{results}. Finally, we present our
conclusions in section \ref{concl}.

\section{The semigrand canonical ensemble} \label{semi_ens}
The most convenient ensemble in the simulation of a
polydisperse system is the so called semigrand canonical
ensemble£¨SCE£© though other ensembles can also be used. In the SCE
the total number of particles $N$  and the volume $V$ are fixed while
the sizes of each particles can be changed. The average particle size
distribution is determined by
the chemical potential difference function $\Delta\mu(\sigma)$.
First, lets consider a system of $N$ hard-spheres in
a volume $V$, and the distribution of the diameter of the spheres
is $P(\sigma)$. Here we assume that the number of particles $N$
is large enough so that the distribution $P(\sigma)$ can be well defined.
The Helmholtz free energy of the system is
\begin{equation}
A=-PV+N\int\mu(\sigma)P(\sigma)d\sigma.
\end{equation}
The semigrand canonical free energy(SCFE) is defined through a
Legendre transform\cite{J. G. Briano}
\begin{equation}
Y=A-N\int(\mu(\sigma)-\mu(\sigma_{r}))P(\sigma)d\sigma.
\end{equation}
Here $\mu(\sigma_{r})$ is the chemical potential of the reference
particle (with diameter $\sigma_r$). SCFE $Y$ is a function of
temperature $T$, volume $V$, total number of particles $N$ and a
functional of $\Delta\mu(\sigma)$. The partition function for SCE is
\begin{eqnarray}
\Gamma &=&\frac{1}{N!}\int_{\sigma_{1}}\cdots\int_{\sigma_{N}}
Z_{N}\left[\prod_{i=1}^{N}
\frac{1}{\Lambda^{3}(\sigma_{i})}\right]\times
\exp\{\beta\sum_{i=1}^{N}(\mu(\sigma_{i})-\mu(\sigma_{r}))\}
\prod_{i=1}^{N}d\sigma_{i}.
\end{eqnarray}
Here $\sigma_i$ and $\Lambda(\sigma_{i})=h/(2\pi m_{i}kT)^{1/2}$ are
the diameter and the thermal wave length of the $i$th  particle,
respectively. And $Z_{N}$ is the canonical configuration integral
\begin{equation}
Z_{N}=\int_{r_{1}}\cdots\int_{r_{N}}\exp(-\beta
U)\prod_{i=1}^{N}d\textbf{r}_{i}.
\end{equation}
By setting
$\mu_{ex}(\sigma_{i})=\mu(\sigma_{i})-kT\ln\left(\frac{N\Lambda(\sigma_{i})^{3}}{V}\right)$
as the excess chemical potential from idea gas, the semigrand
canonical partition function can be written in a more symmetrical
form
\begin{eqnarray}
\Gamma
=\frac{1}{N!\Lambda^{3N}(\sigma_{r})}\int_{\sigma_{1}}\cdots\int_{\sigma_{N}}
Z_{N}\times
\exp\{\beta\sum_{i=1}^{N}(\mu_{ex}(\sigma_{i})-\mu_{ex}(\sigma_{r}))\}
\prod_{i=1}^{N}d\sigma_{i}.
\end{eqnarray}
The SCFE of the system is related to  the partition function by the
following relation
\begin{equation}
Y=-kT\ln\Gamma(N,V,T,{\Delta\mu(\sigma)}).
\end{equation}
Thus the stable state can be obtained in the semicanonical ensemble
by the minimization of the semicanonical free energy, which is the
criteria for the stability of the polydisperse hard sphere crystal.

In practical simulation calculations, the diameter of particles is
discretised and the corresponding  semigrand canonical partition
function is
\begin{equation}
\Gamma=\frac{1}{N!\Lambda^{3N}(\sigma_{r})}\sum_{\sigma_{L}=\sigma_{s}}^{\sigma_{L}}\cdots
\sum_{\sigma_{N}=\sigma_{s}}^{\sigma_{L}} Z_{N}\times
\exp\{\beta\sum_{i=1}^{N}(\mu_{ex}(\sigma_{i})-\mu_{ex}(\sigma_{r}))\},
\end{equation}
where $\sigma_{s}$ and $\sigma_{L}$are the maximum and minimum
values of the particle diameters, respectively. The above discussion
can be extended straight forwardly to the polydispersity of other
properties of the particles, such as charge dispersity, shape
dispersity and mass dispersity etc.

\section{The methods} \label{method}

\subsection{NEPR}
Under the semigrand canonical ensemble (SCE), the  excess chemical
potentials for different particles (different diameters) relative to
the reference particle(diameter $\sigma_{r}$),
$\mu_{ex}(\sigma_{i})-\mu_{ex}(\sigma_{r})$, are given. However, in
experiments the fixed quantity is the distribution of particle
diameters $P(\sigma)$, in order to simulate the experimental
controllable system with given particle size distribution, a proper
excess chemical potential has to be chosen which can reproduce the
required particle size distribution. This is in fact the solution of
the functional equation
$\Delta\mu_{ex}(\sigma)=\Delta\mu_{ex}(\{P(\sigma)\})$. Wilding has
proposed an effective and robust procedure, the nonequilibrium
potential refinement (NEPR) algorithm\cite{N. B. Wilding1},
 to tackle this problem. The algorithm can be used to solve a wide range of the
 so-called inverse problems\cite{D. Levesque} such as obtaining the inter particle
 interactions from experiment measured structure factors. It can also be used
 in our problem to find the excess chemical potential from
 the particle size distribution. The original NEPR  algorithm was developed in the
 framework of the grand canonical ensemble, we extended it to the case of the SCE and
 used it in the calculation reported here. The following is the detailed description
 of the extension.

Consider a polydisperse system of hard spheres, the range of the
particle diameter is $\sigma_{s}=\sigma_{1}\cdots
\sigma_{i}\cdots\sigma_{c}=\sigma_{L}$, and the diameter of
particles can take $c$ discrete values, $\sigma_i$, $i=1$, $2$,
 $\cdots$, $c$.
When $c$ is large enough, the diameter of the particles tends to a continuous variable
which can resemble the real polydisperse system. The diameter distribution is
 $P(\sigma)$, which is normalized in the following way
\begin{equation}
\sum_{i=1}^{c} P(\sigma_{i})=1.
\end{equation}

The excess chemical potential  $\Delta\mu_{ex}(\sigma)$ is solved by
simulation in a recurrence way. First, initial guess of the excess
chemical potential is assigned, then it is modified at every few
Monte Carlo steps according to the instant diameter distribution,
$P_{ins}(\sigma)$, which records the distribution of particles {at
the instant} of the simulation, the simulation is terminated when
the average of $P_{ins}(\sigma)$ is the same as the required
distribution $P(\sigma)$ within some tolerance. Then the
$\Delta\mu_{ex}(\sigma)$ is the solution of the problem. The initial
value of the excess chemical potential is not a vital factor in the
calculation process and may be assigned any reasonable values, for
example, $\Delta\mu_{ex}(\sigma)=1$ for all diameters. The detail
implement is the following.
\subsubsection{The particle move}
There are two kinds of particle moves in the simulation, the first
kind is the random displacement of a randomly chosen particle, which
is rejected  or accepted depends on whether the new position
overlaps to other particles or not, the second kind is the expansion
or retraction of a randomly chosen particle, which is named as
breathing move in literature\cite{M. R. Stapleton}, the probability
of acceptance of a breathing move which does not result in an
overlap is
\begin{equation}
P_{acc}=\min\left\{1,\exp\{\beta(\Delta\mu_{ex}(\sigma^{'}_{i})-
\Delta\mu_{ex}(\sigma_{i}))\}\right\},
\end{equation}
where $\sigma_{i}$ and $\sigma^{'}_{i}$ are diameters of the $i$th
particle before and after the test move. The move is rejected if it
results in an overlap with other particles.

\subsubsection{The iteration}
For a given particle size distribution, the excess chemical
potential is calculated  by a Monte Carlo iteration procedure. The
central quantity in this procedure is the instantaneous particle
size distribution $P_{ins}(\sigma)$, which is the histogram of the
particle size distribution at the instant of the simulation and
updated during the simulation. Another important quantity is the
average particle size distribution $\overline{P}(\sigma)$, which is
the average of the instantaneous particle size distribution in the
simulation. The excess chemical potential is updated by the
Wilding's scheme\cite{N. B. Wilding1} for every short intervals. The
Wilding's scheme in this iteration is given by
\begin{equation}
\Delta\mu^{'}_{ex}(\sigma)=\Delta\mu_{ex}(\sigma)-
\gamma_{i}\left(\frac{P_{ins}(\sigma)-P(\sigma)}{P_{ins}(\sigma)}\right)\qquad \forall
\sigma.
\end{equation}
Here $P(\sigma)$ is the given particle size distribution,
$\gamma_{i}$ is a modification factor of the $i$th iteration.
For a given modification factor, the average size distribution
$\overline{P}(\sigma)$ is also recorded during the
simulation. When the difference of the average size distribution and
the given particel size distribution is less than a specified  value $\xi$
 \begin{equation}
\xi\geq \max\left(\left|\frac{\overline{P}(\sigma)-P(\sigma)}{P(\sigma)}\right|\right),
\end{equation}
one loop of the iteration is finished. The modification factor is
then reduced by a factor $1/n$ where $n$  is a small integer,
$\gamma_{i+1}=\gamma_{i}/n$, and the excess chemical potential of
the last iteration is used as the initial input and start the next
iteration. The iteration continues till  the modification factor
$\gamma$ reaches a very small value, and the resulted excess
chemical potential is then regarded as the solution of the problem.
In practical calculations,  the convergent  criteria for
$\gamma$ , the threshold $\xi$ and the reduced factor $n$ are tuned
to reach both high  efficiency and accuracy.

The NEPR algorithm for SCE was tested by the simulation of a
polydisperse hard sphere fluid and a polydisperse hard sphere
crystal with $fcc$ structure. In the tests the particle
size distribution is chosen to be the Schultz distribution, which
is the most studied  model distribution in polydisperse
systems. The distribution is
\begin{equation}
P(\sigma)=\frac{1}{z!}\left(\frac{z+1}{\overline{\sigma}}\right)^{z+1}\sigma^{z}
\exp\left[-\left(\frac{z+1}{\overline\sigma}\right)\sigma\right],
\end{equation}
where $\overline{\sigma}$ is the average diameter of the particles
and $z$ controls the width of the distribution. In the Schultz
distribution, the range of the diameter is $[0,+\infty)$, however,
in a simulation calculation with finite number of particles, cuts
off of the up and the lower limit of the distribution may be
specified for  convenience of computation. In the test studies
the effect of the cutoff is not studied, the emphases is on the
effect of polydispersity to the physical properties of the system.
On the other hand, small particles may enter into the interstitial
space of crystals and induce instabilities of crystal structure,
this is beyond the subject of this study though it is an interesting
subject of research.

In the test simulation, the threshold   $\xi=0.15$, the initial
modification factor $\gamma_{0}=0.01$ and the reduce factor is
$1/2$. The termination criteria is $\gamma\leq0.0001$. For the hard
sphere liquid, the volume fraction $\phi=0.3$ and the dispersity
$\delta=14.2\%$; for the $fcc$ hard sphere crystal, the volume
fraction $\phi=0.6$ and the dispersity   $\delta=3.8\%$. Figure
\ref{fig1} and figure \ref{fig2} are the simulation results. Figure
\ref{fig1}(a) is the calculated excess chemical potential for the
truncated Schultz distribution as function of the diameter of
particles in the liquid state,  figure \ref{fig1}(b) is the
comparison of the given truncated Schultz distribution and the
distribution generated with calculated excess chemical potential,
the agreement between the two is excellent. Figure \ref{fig2}(a) and
figure \ref{fig2}(b) are the calculated excess chemical potential
and the comparison of given and generated distributions in the
crystal state, respectively, the agreement is also excellent as in
the liquid case.

\subsection{Lattice switch}

The free energy difference between the $fcc$ and the $hcp$ hard  sphere
crystals is extremely small, in order to obtain a reliable result
for the difference, we need to find an accuracy method of
calculation. There are different methods suggested in the past in
the studies of monodisperse hard sphere crystals and many results
were obtained for the problem.  The lattice switch
method(LW)\cite{A. D. Bruce,A. D. Bruce1} developed recently is a
high precession method in the calculation of the free energy
difference. The method has extended successfully to the calculation
of the liquid-solid transition of monodisperse hard sphere
systems\cite{N. B. Wilding2} and also used in the studies of soft
sphere systems\cite{A. N. Jackson,J. R. Errington}. In this
subsection, we extend it to the SCE.

A detailed presentation of the LW method for monodisperse hard
sphere crystals in the canonical ensemble can be found in references
\cite{A. D. Bruce,A. D. Bruce1}. Here we give a quick sketch of the
method in the context of polydisperse hard sphere crystals. The
system contains  $N$ hard spheres in volume $V$ with periodic
boundary conditions. The hard spheres can be in the $fcc$ or the
$hcp$ structures respectively, and the spatial positions of the
particles are specified by the position vectors
$\mathbf{r}_{fcc i}$ or $\mathbf{r}_{hcp i}$ for the
$i$th particle in the $fcc$ structure or $hcp$ structure,
respectively. The position vectors can be decomposed as
\begin{equation}
\mathbf{r}_{\alpha i}=\mathbf{R}_{\alpha i}+\mathbf{u}_{\alpha i},
\end{equation}
where the subscript $\alpha$ may represent $fcc$ or $hcp$. The
$\mathbf{R}_{fcci}$ and $\mathbf{R}_{hcpi}$ are
lattice vectors of the idea structures, $\mathbf{u}_{fcc
i}$ and $\mathbf{u}_{hcp i}$ are the displacements from the
idea structures.

In principle, the displacement can be any vectors that only
constrained by the geometry of the simulation box, however, in the
crystal phase with dispersity smaller than the terminal dispersity,
the displacements are naturally cutoff in the simulation time scale.
We use $\{\mathbf{u},\mathbf{\sigma}\}$ to represent
the phase space of the polydisperse system, here
$\mathbf{\sigma}$ denotes the diameters of $N$ particles.
Each structure $\alpha$($\alpha= fcc \hbox{ or } hcp$)
 associates a set of displacements
 $\{\mathbf{u},\mathbf{\sigma}\}_{\alpha}$.
 In a typical simulation, in which a representative subset of the
 displacements for one structure is sampled, the transition from one structure to another
 can not happen because the transition probability between structures
is extremely small. The spirit of the lattice switch method is that
switch the ideal lattice vectors from one structure to another while
keep the displacements frozen. The two sets
$\{\mathbf{u},\mathbf{\sigma}\}_{fcc}$ and
$\{\mathbf{u},\mathbf{\sigma}\}_{hcp}$ have a common intersection
$\{\mathbf{u},\mathbf{\sigma}\}_{fcc}\bigcap
\{\mathbf{u},\mathbf{\sigma}\}_{hcp}$, which provides a gate to
relate  the two structures. All allowed(non overlap) configurations
accessible by simulation which are associated with $fcc$ and $hcp$
structures can be divided into three subsets, (a). all  the
displacements allowed by the $fcc$ structure but not allowed by the
$hcp$ structure, which we  denote as
$\{\mathbf{u},\mathbf{\sigma}\}_{fcc-hcp}$, (b). all the
displacements allowed by the $hcp$ structure but not allowed by the
$fcc$ structure, which we denote as
$\{\mathbf{u},\mathbf{\sigma}\}_{hcp-fcc}$, (c). the displacements
allowed by both the $fcc$ structure and the $hcp$ structure, denoted
as $\{\mathbf{u},\mathbf{\sigma}\}_{fcc\bigcap hcp}$.

The semigrand canonical ensemble partition function of the two structures can
be written as
\begin{eqnarray}
\Gamma(\alpha)
=\frac{1}{N!\Lambda^{3N}(\sigma_{r})}\int_{\mathbf{\sigma}\in
\{\mathbf{\sigma}\}}
\prod_{i=1}^{N}d\sigma_{i}\int_{\mathbf{u}\in
\{\mathbf{u}\}} \prod_{i=1}^{N} d\mathbf{u}_{i}
\times
\exp\{\beta(\sum_{i=1}^{N}\Delta\mu_{ex}(\sigma_{i})-\phi(\mathbf{u},\alpha))\},
\end{eqnarray}
where $\phi$ is the potential energy of the system, which is
$\infty$ or $0$ for the hard sphere system. The SCFE for the
structure $\alpha$ is
\begin{equation}
Y(\alpha)=-kT\ln \Gamma(\alpha).
\end{equation}
The SCFE difference of the two structures can be written as
\begin{eqnarray}
Y_{hcp}-Y_{fcc}&=&kT\ln \frac{\Gamma(fcc)}{\Gamma(hcp)} \notag
\\
&=&kT\ln\frac{P_{fcc}}{P_{hcp}} \notag
\\
&=&kT\ln\frac{P_{fcc-hcp}+P_{fcc\bigcap
hcp}}{P_{hcp-fcc}+P_{fcc\bigcap hcp}}.
\end{eqnarray}
Where $P_{fcc-hcp}$\,,  $P_{hcp-fcc}$ and $P_{fcc\bigcap hcp}$
represent the probabilities of three subsets
$\{\mathbf{u},\mathbf{\sigma}\}_{fcc-hcp}$,
$\{\mathbf{u},\mathbf{\sigma}\}_{hcp-fcc}$ and
$\{\mathbf{u},\mathbf{\sigma}\}_{fcc\bigcap hcp}$
respectively.

It is clear from the above discussion that the calculation of
semigrand free energy difference from simulation may be proceed as
follows: first, the excess chemical potentials for both structures
are determined by the iteration procedure described in the last section.
Then starting from any structure, particle size distribution
 and displacement, regard
all of the displacement, particle size and the structure index as
random variables, and make Monte Carlo moves. In principle, the
system will move among the $fcc-hcp$, $hcp-fcc$ and $fcc\bigcap hcp$
states, by recording the number of microstates corresponding to the
three macrostates, the probabilities of each macrostate can be
obtained and then follows the free energy difference. However, this
prescription is not practical in real simulations, the system will
trap in either $fcc-hcp$ or $hcp-fcc$ macrostate in the simulation
period simply because the number of microstates of $fcc-hcp$ and
$hcp-fcc$ are much larger than the number of microstates of
$fcc\bigcap hcp$. In order to overcome this difficulty, the bias
sampling can be used. To achieve this goal, we first define an order
parameter $\mathscr{M}(\mathbf{u},\mathbf{\sigma})$ for the
displacement field,
\begin{equation}
\mathscr{M}(\mathbf{u},\mathbf{\sigma})=M(\mathbf{u},\mathbf{\sigma},hcp)
-M(\mathbf{u},\mathbf{\sigma},fcc).
\end{equation}
Here $M(\mathbf{u},\mathbf{\sigma},hcp)$ and
$M(\mathbf{u},\mathbf{\sigma},fcc)$ represent the
number of overlap pairs of the $hcp$ and $fcc$ structure for all
samples of the displacements. The order parameter $\mathscr{M}$ can
take values of $0$, $\pm1$, $\pm2$, $\cdots$, where $\mathscr{M}=0$
corresponding to the macrostate $fcc\bigcap hcp$. In the simulation
process one of the
$M(\mathbf{u},\mathbf{\sigma},hcp)$ and
$M(\mathbf{u},\mathbf{\sigma},fcc)$ has to be zero
since the domain of random walk is
$\{\mathbf{u},\mathbf{\sigma}\}_{fcc}\bigcup
\{\mathbf{u},\mathbf{\sigma}\}_{hcp}$. The free
energy difference can be represented by the macroscopic order
parameter $\mathscr{M}$ as
\begin{equation}
Y_{hcp}-Y_{fcc}=kT\ln\frac{\sum_{\mathscr{M}\geq0}P(\mathscr{M})}
{\sum_{\mathscr{M}\leq0}P(\mathscr{M})},
\end{equation}
here $P(\mathscr{M})$ is the probability that the order parameter
takes value $\mathscr{M}$. Now we regard each value of the order
parameter corresponding to a macroscopic state, biased sample
alorithm\cite{B. A. Berg,F. wang,J. S. Wang} is to sample the system
according to a probability so that the rate of visits to every
macrostate is basically the same. If the sampling probability is the
inverse of $P(\mathscr{M})$, then the rate of visits to each
macrostate is exactly the same. Unfortunately, $P(\mathscr{M})$ is
the quantity we are looking for which is unknown before calculation.
The problem was solved by  several different methods in the  context
of density of states calculations, from which  the multicanonical
method\cite{B. A. Berg} and Wang-Landau method\cite{F. wang} are the
most used. These methods, when used in the current problem,  amount
to starting the calculation with an initial guess of the probability
$P(\mathscr{M})$,  sampling the system with inverse of this
probability, and modify $P(\mathscr{M})$ according to the visit
rates to the macrostates till the visit rates are constant for  all
macrostates within a given tolerance, then the resulted
$P(\mathscr{M})$ and  visit rates  together will give an accurate
estimate of the free energy difference.  One of the implementaion is
to introduce a weight function $\eta(\mathscr{M})$, sampling  the
system with
$e^{\beta(\sum_i\Delta\mu_{ex}(\sigma_i)-\phi(\mathbf{u}))+\eta(\mathscr{M})}$,
calculating the probability distribution  of the order parameter
$\mathscr{M}$, $P_{\eta}(\mathscr{M})$, modifying the  weight
function $\eta(\mathscr{M})$ till  $P_{\eta}(\mathscr{M})$ is close
to constant and then obtain the required probability
$P(\mathscr{M})$ through
\begin{equation}
P(\mathscr{M})=P_{\eta}(\mathscr{M})e^{-\eta(\mathscr{M})}.
\end{equation}
Based on these discussions, the acceptance probability of the
sampling is summarized in the following expression
\begin{equation}
P_{acc}(\mathscr{M},\mathbf{u},\mathbf{\sigma}\rightarrow
\mathscr{M}^{'},\mathbf{u}^{'},\mathbf{\sigma}^{'})=
\min\left\{1,\frac{\exp(\beta(\sum_{i}\Delta\mu_{ex}(\sigma^{'}_{i})-\phi(\mathbf{u}^{'}))
+\eta(\mathscr{M}^{'}))}
{\exp(\beta(\sum_{i}\Delta\mu_{ex}(\sigma_{i})-\phi(\mathbf{u}))+
\eta(\mathscr{M}))} \right\},
\end{equation}
where$(\mathscr{M},\mathbf{u},\mathbf{\sigma})$ and
$(\mathscr{M}^{'},\mathbf{u}^{'},\mathbf{\sigma}^{'})$
are the order parameters, displacement fields and diameters of the system
 corresponding to states before and after
the test move, respectively.

\section{Computational details and Results}  \label{results}

In this section we use the extended NEPR and lattice switch method
described in the last section to study the stability of the
polydisperse hard sphere crystal. The distribution of the hard
spheres is chosen to be the Schultz distribution as was used by many
researchers. The chemical potential difference obtained from this
distribution with $fcc$ lattice can reproduce fairly accurate
Schultz distribution for the specified  $hcp$ lattice, this is
because of the difference between the two structures is very
small(see Fig. \ref{fig5}).
 In fact, if we specify a chemical potential
difference, we may produce the same particle size distributions in
both of the $fcc$ and the $hcp$ structures within the statistical
errors. It is well known that there is a terminal poly-dispersity in
polydisperse crystals above which the bulk crystal may not stably
existed. The terminal polydispersity $\delta_{t}$ obtained from
recent simulation is about $5\sim6\%$\cite{P.G. Bolhuis,D. A.
Kofke}. In our simulation we set the maximum dispersity to be $4\%$
so that the stable crystal can be simulated. The simulation boxes
are set up to suit the idea $fcc$ and $hcp$ crystal, periodic
boundary conditions are used in the calculation, the initial
configuration are idea $fcc$ and $hcp$ lattices, respectively. The
Wang-Landau sampling method was used to obtain a crude estimation of
the weight function of the order parameter, then the multicanonical
algorithm is used to refine the result. The calculated results are
shown in Table \ref{table1}, for the polydispersity used in the
simulation, the $fcc$ lattice has the smaller free energy and more
stable than the $hcp$ crystal. In the case of $\delta = 4\%$, there
are possibilities that the smallest sphere may jump from the cage of
its lattice positions so that a defect may be created,  to avoid
this situation we  used a larger volume fraction as shown in the
last column of Table \ref{table1}. The effect of finite size effect
was studied in the case of medium polydispersity by enlarging the
simulation boxes with fixed volume fraction, it was found that the
value of the free energy is affected by the box size slightly but
the relative stability is unchanged.

In order to have a clear picture of the relative stability of the
structures, we plotted the probability distribution of the
macrostates  for two different polydispersities in figures
\ref{fig6}--\ref{fig8}. For convenience of comparison, we plotted
the distribution of $fcc$ and $hcp$ in the same half plane, in the
$hcp$ case  the absolute value of the order parameter is used. From
the figures we see that there is a maximum of probability for each
structure, and the probability maximum of $fcc$ structure is larger
then that of the  $hcp$ structure which means that $fcc$ structure
is more favorable. Figure \ref{fig9} shows that the ratio of the
probability of ``gate'' states ($\mathscr{M}=0$) to the probability
maximum is about $10^{-35}$ which means that the  ``gate'' states
will never be reached if a simple sampling scheme is used.
Considering the extreme difference of the probability between the
``gate'' state and the maximum state, the refinement of macrostates
and bias sampling are the necessary scheme to obtain meaningful
results.

The particle size  distribution used in the calculation is the
truncated schultz distribution, this is the widely used distribution
in theoretical studies. The results based on this distribution may
not be extended to all polydisperse systems, however, we expect that
it represents a class of polydisperse systems. The distribution
dependence of structures of polydisperse systems requires detailed
computations of various polydisperse systems. It should also be
noted that the accuracy of our calculation is limited by the
computation resources, the  number of particles is pretty small,
typical Monte Carlo steps are $50$ million which is still too small
to determine accurately the relationship between the difference of
semigrand canonical free energy and the polydispersity. On the other
hand, the conclusion that the $fcc$ is more favorable than the $hcp$
structure for the polydisperse hard sphere crystals with truncated
Schultz distribution of diameters of spheres is clear and reliable.

\section{conclusion and discussion} \label{concl}
Polydispersity of colloid system is common in artificial colloids.
We studied here the effect of polydispersity on the stability of
structures of colloid crystals, and found that $fcc$ structure is
more stable than the $hcp$ structure, which is the same as the
monodisperse case with the same calculations. To study the problem
we have extended the NEPR algorithm and the lattice switch method to
the semigrand ensemble. The extension provides a powerful tool in
the studies of other thermodynamical problems of polydisperse
systems. A direct application is the determination of the phase
diagrams of the polydisperse systems which may replace or at least
complement the current Gibbs-Duhem integration method\cite{P.G.
Bolhuis,D. A. Kofke}. The monodisperse system has already studied by
the original LW method\cite{N. B. Wilding2}. The above extension can
also be extended to the problems of soft sphere system simulations
by some extra techniques\cite{A. N. Jackson,J. R. Errington}.

\begin{acknowledgments}
 The
work is supported by the National Natural Science Foundation of
China under grant No.10334020 and in part by the National Minister
of Education Program for Changjiang Scholars and Innovative Research
Team in University.
\end{acknowledgments}

\newpage
\begin{table}[http]
\begin{tabular}{|c|c|c|c|c|c|c|}
  \hline
  $\phi$ & 0.576 &0.576 &0.576 & 0.576 & 0.576 & 0.602 \\\hline
  $\delta$ & 0 & 0&1.1$\%$ & 2.5$\%$ & 2.5$\%$ & 4$\%$ \\\hline
  N & 216 &1728 & 216 & 216 & 1728 & 216 \\\hline
  $\Delta f\times10^{5}$ & 133(3)& 113(3)& 139(11) & 133(16) & 110(15) & 170(20) \\
  \hline
\end{tabular}
\caption{Parameters and simulation results: $\phi$ is the volume
fraction of the system, $\delta$ is the polydispersity of particle
diameters (same for both $fcc$ and $hcp$ structures),
 $N$ is the total number of particles of the system, $\Delta
f=(Y_{hcp}-Y_{fcc})/N$ is the semigrand canonical free energy
difference between $hcp$ and $fcc$ structures, in unit of $k T$. The
numbers in the parentheses is the uncertainty of the result. The
results of monodisperse hard sphere system(from reference \cite{A.
D. Bruce1}) are listed in the first two columns for comparison.
}\label{table1}
\end{table}

\newpage

\begin{spacing}{2.0}

FIG. 1:(a), The solved excess chemical potential difference of hard
spheres as function of particle diameter $\sigma$ in the fluid
state. (b), The line is the plot of the Schultz function, and the
dots are the particle diameter distribution obtained from simulation
by using the $\Delta\mu_{ex}(\sigma)$ plotted in (a).
\\

\begin{figure}[http]
\centering \subfigure[]{
\label{fig1:subfig:a} 
\includegraphics[width=3.2in]{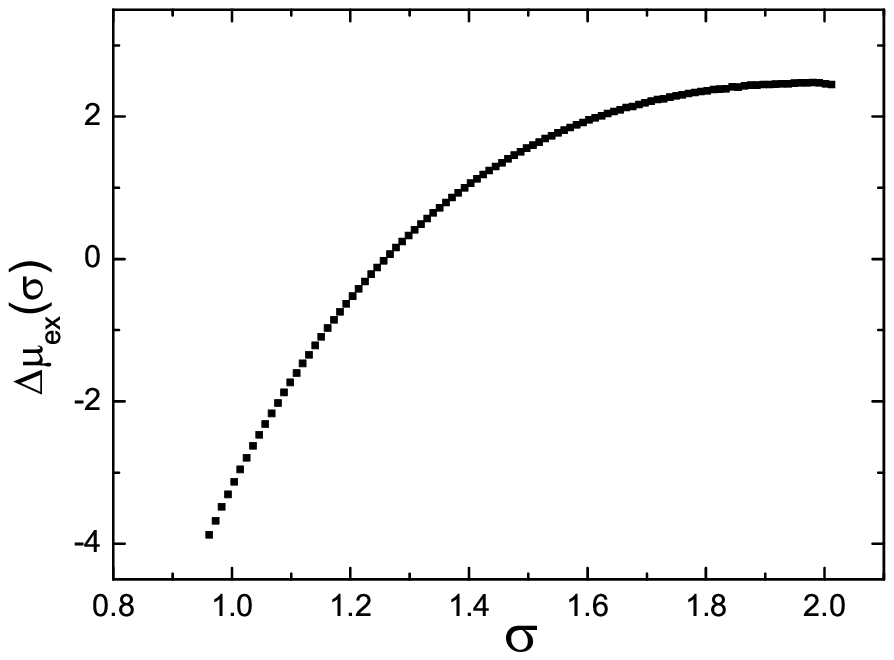}}
\hspace{-0.3in} \subfigure[]{
\label{fig1:subfig:b} 
\includegraphics[width=3.2in]{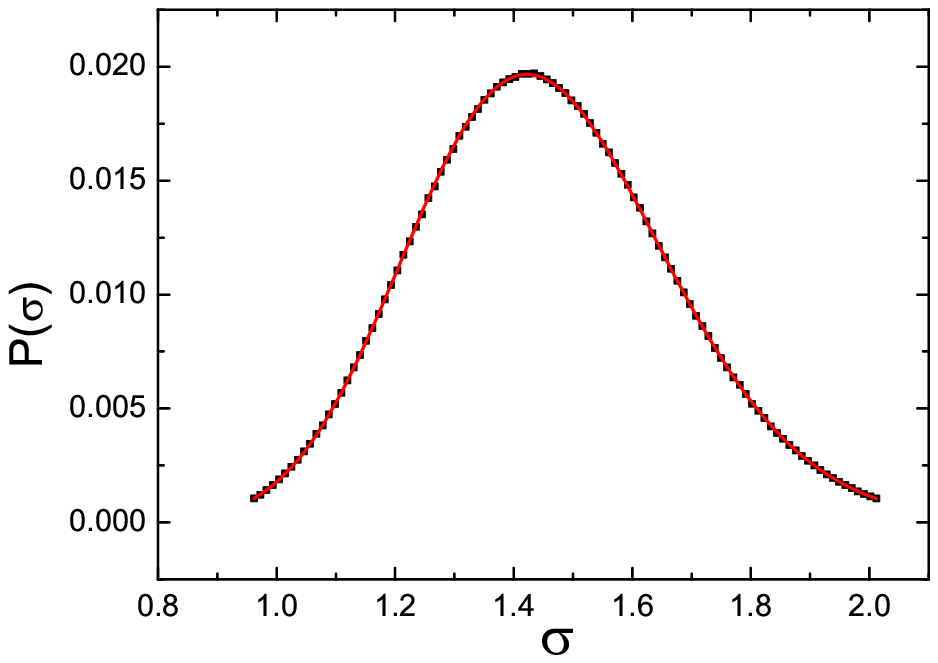}}
\caption{}
\label{fig1} 
\end{figure}

\begin{figure}[http]
\centering \subfigure[]{
\label{fig2:subfig:a} 
\includegraphics[width=3.2in]{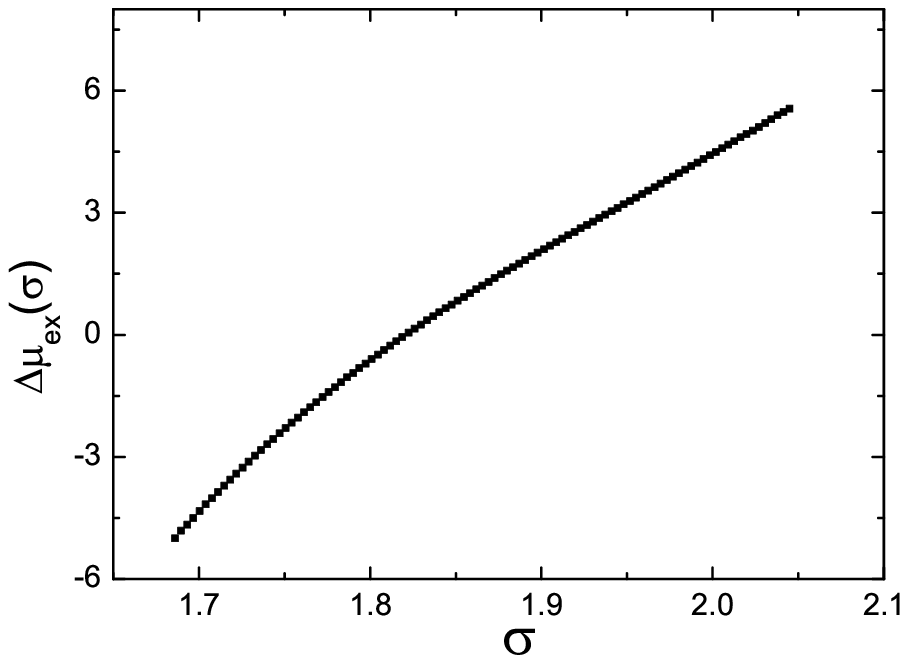}}
\hspace{-0.3in} \subfigure[]{
\label{fig2:subfig:b} 
\includegraphics[width=3.2in]{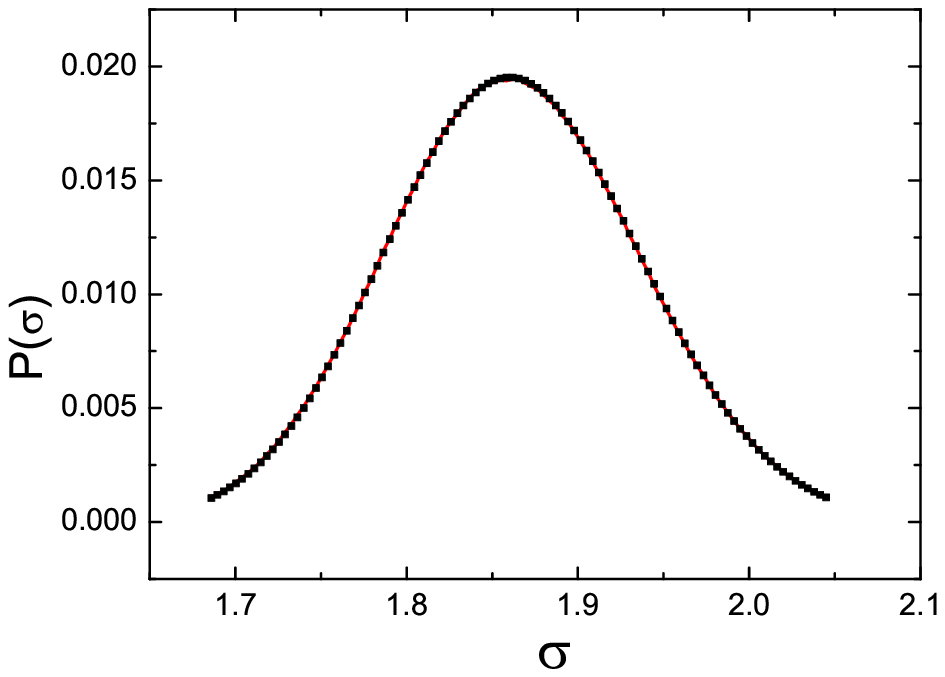}}
\caption{}
\label{fig2} 
\end{figure}

\begin{figure}[http]
\includegraphics[angle=0,width=0.8\textwidth]{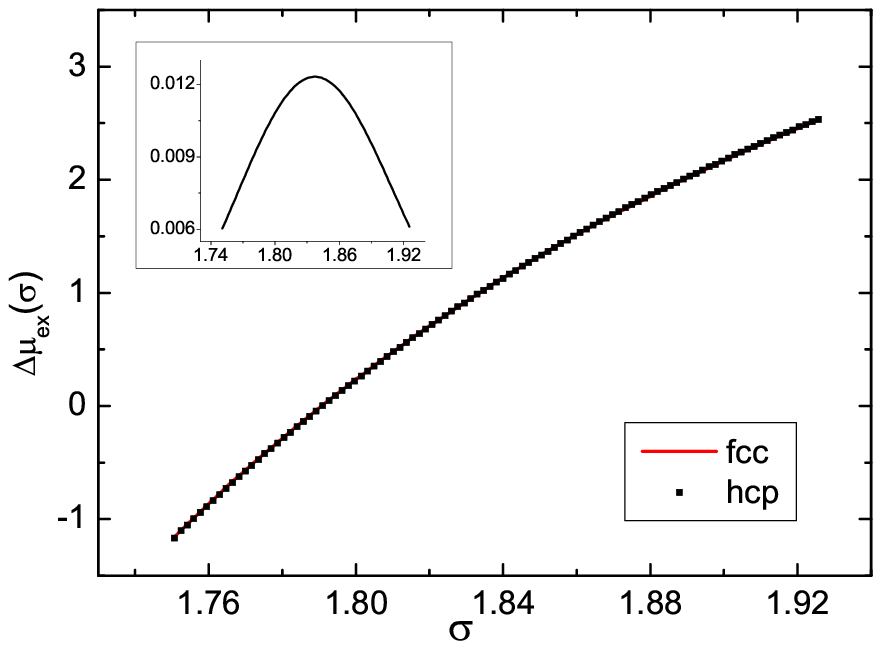}
\caption{}
\label{fig5}
\end{figure}

FIG. 2: The same as figure \ref{fig1}, for $fcc$ solids.
\\

FIG. 3: The calculated excess chemical potentials for the truncated
Schultz distribution of particle diameters from NEPR method. The
line is for the $fcc$ structure and the points are for the $hcp$
structure, the difference is smaller then the statistical errors.
\\

FIG. 4: Left: the excess chemical potential as function of the
particle diameter, insert is the particle size distribution for
$fcc$ crystal. Right:the distribution of macrostates (order
parameters) for $fcc$ (open circles) and $hcp$ (filled square)
crystals. The polydispersity $\delta=1.1\%$, volume fraction
$\phi=0.576$.
\\

\begin{figure}[http]
\includegraphics[angle=0,width=0.5\textwidth]{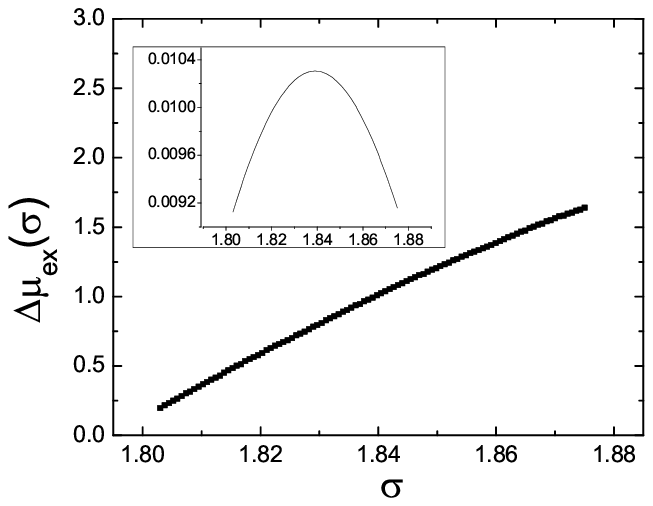}
\includegraphics[angle=0,width=0.48\textwidth]{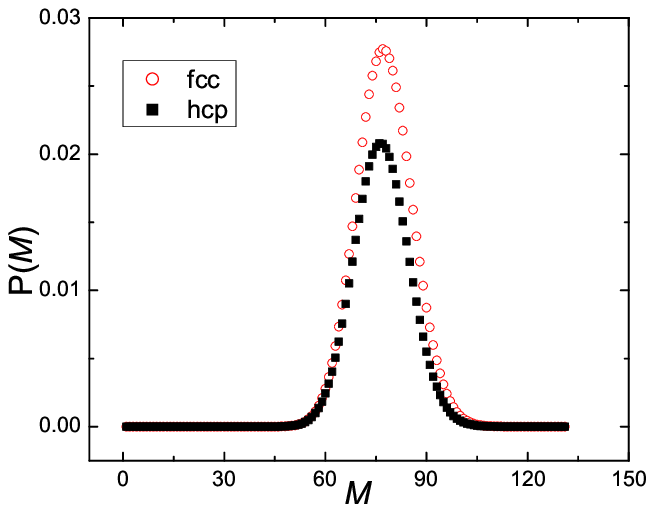}
\caption{}
\label{fig6}
\end{figure}

\begin{figure}[http]
\includegraphics[angle=0,width=0.49\textwidth]{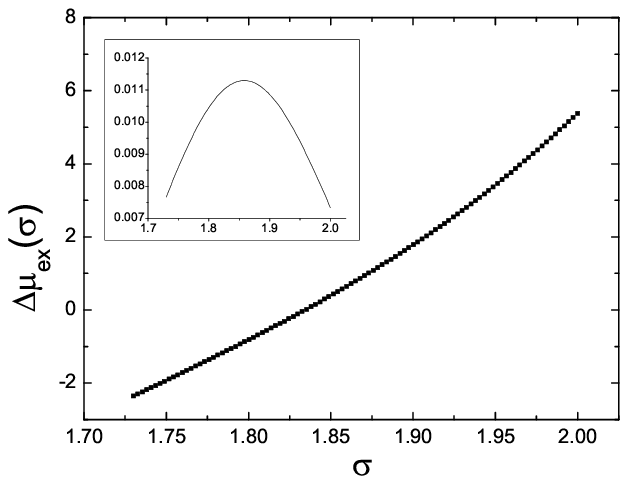}
\includegraphics[angle=0,width=0.49\textwidth]{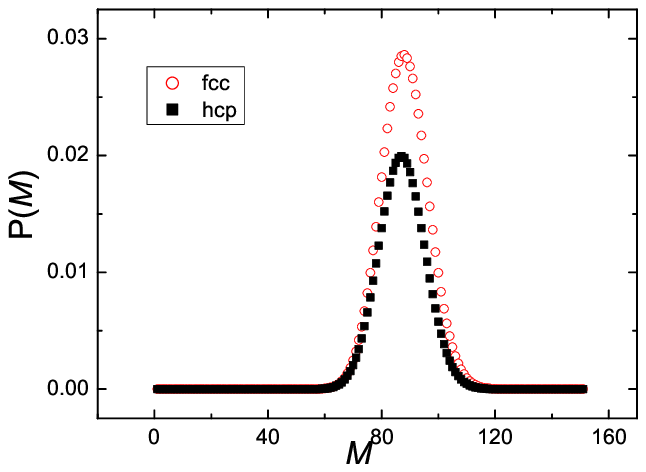}
\caption{}
\label{fig8}
\end{figure}

\begin{figure}[http]
\includegraphics[angle=0,width=0.8\textwidth]{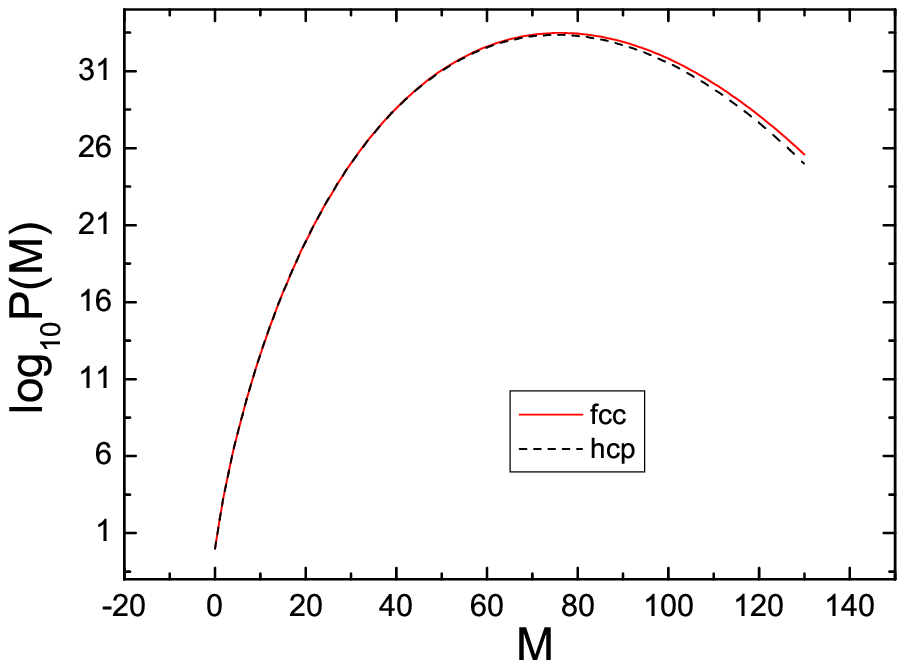}
\caption{}
\label{fig9}
\end{figure}

FIG. 5:  Same as figure \ref{fig6}, $\delta=4\%$, volume fraction
$\phi=0.602$.
\\

FIG. 6: The logarithm of the probability distribution of macrostates
of hard sphere crystals, the solid line is for the $fcc$ crystal and
the dashed line is for the $hcp$ crystal.

\end{spacing}

\newpage

\newpage

\newpage

\newpage

\newpage

\newpage

\end{document}